\documentclass[prapplied, amsmath,amssymbn,twocolumn,superscriptaddress]{revtex4-1}
\usepackage{graphicx}
\usepackage{color}
\usepackage[utf8]{inputenc}

\usepackage{blindtext}
\usepackage{chemformula}

\begin{document}

\title{Out-of-equilibrium Polymorph Selection in Nanoparticle Freezing}

\author{Jonathan Amodeo}
\affiliation{Université de Lyon, INSA-Lyon, MATEIS, UMR 5510 CNRS, 69621 Villeurbanne, France}
\author{Fabio Pietrucci} 
\affiliation{Sorbonne Université, CNRS UMR 7590, IMPMC, 75005 Paris, France}
\author{Julien Lam}
\email{julien.lam.pro@gmail.com}
\affiliation{Center for Nonlinear Phenomena and Complex Systems, Universit\'{e} Libre de Bruxelles, Code Postal 231, Boulevard du Triomphe, 1050 Brussels, Belgium}


\makeatletter
\newcommand*{\forcekeywords}{
  \acs@keywords@print
  \let\acs@keywords@print\relax
}
\makeatother

\begin{abstract}
The ability to design synthesis processes that are out of equilibrium has opened the possibility of creating nanomaterials with remarkable physico-chemical properties, choosing from a much richer palette of possible atomic architectures compared to equilibrium processes in extended systems. In this work, we employ atomistic simulations to demonstrate how to control polymorph selection via cooling rate during nanoparticle freezing in the case of Ni$_3$Al, a material with a rich structural landscape. State-of-the-art free-energy calculations allow to rationalize the complex nucleation process, discovering a switch between two kinetic pathways, yielding the equilibrium structure at room temperature and an alternative metastable one at higher temperature. 
Our findings address the key challenge in the synthesis of nano-alloys for technological applications, i.e., rationally exploiting the competition between kinetics and thermodynamics by designing a treatment history that forces the system into desirable metastable states. 
\end{abstract}

\maketitle

Crystallization is a complex and incompletely understood phenomenon that has long fascinated scientists from various research fields including environmental science\cite{Campbell2018Apr,Myint2018Oct,Tabazadeh2002Dec,Pipolo2017Dec}, pharmaceutics\cite{Ma2017Apr,Varongchayakul2013Sep} and materials engineering\cite{Tidhar2014Sep,Shen2015Oct,Fei2019Jan}. One of the key challenges of crystallization remains the  ability to control the polymorph selection since properties of most materials among which mechanical, optical and chemical depend drastically on the crystal structure. This intricate relationship between structure and properties is particularly important in nanomaterials where novel crystal phases can be stabilized thus enabling for unique technological applications\cite{Nguyen2015Dec,Sharma2019Apr}.

Although the understanding of polymorph selection at the bulk scale has been the subject of numerous works\cite{Desgranges2006Nov,Desgranges2007Feb,Desgranges2006Aug,Russo2012Jul,Russo2012}, it remains unclear to what extent the specific features of nanoparticles, including finite-size effects and surface preponderance, can affect the resulting structure. As far as nanoparticles are concerned, many simulations studies have focused on finding minimum energy configurations\cite{Laurens2020Jan,Hong2015Oct,Morales-Garcia2019May,Lam2015Aug,MaciaEscatllar2019Jun,Vines2017Jul} and on studying the nucleation path near the melting temperature\cite{Desgranges2009Mar,Quigley2008Jun,Quigley2011Jan,Ngale2009Dec,Ou2019Oct,Pavan2015Nov}. 

However, from the experimental point of view, nanoparticles are usually obtained through out of equilibrium processes especially when using physical routes of synthesis such as laser ablation~\cite{Ayyub2001Jul,Semaltianos2010May,Zhang2017Feb,Kim2017}, flame pyrolysis~\cite{Rosnerc2005Jul,Teoh2010Aug,Li2016Jul}, and magnetron sputtering~\cite{Ayyub2001Jul,Kelly2000Mar,Alexeeva2016Feb}. In all these cases, as the temperature is decreased, the system goes from the gas phase to a liquid droplet whose subsequent freezing lead to the formation of a nanocrystal\cite{Kesala2007May,Zhao2015Jan,Forster2019Oct}. While being crucial to reach a rational design of nanoparticles, the highly complex nucleation pathways and the resulting polymorph selection remain difficult to understand in this out of equilibrium context. In particular, numerous works investigated the freezing mechanisms in nanoparticles\cite{Nam2002Dec,Qi2008May,Lewis1997Jul} and how the size and the chemical composition of the liquid droplet can influence the final structure\cite{Chushak2003Apr}. However, the cooling rate -- an important tuning parameter in synthesis experiments -- has been scarcely studied\cite{Rossi2018Jun,Kesala2007May,Forster2019Oct,Qi2008May} even though it provides a primary route to study the interplay between kinetic and thermodynamical pathways. 

Our work investigates how the cooling rate controls polymorph selection that occurs during nanoparticle freezing. We take the example of \ch{Ni_3Al} which is an intermetallic alloy. At the nanometric scale, this class of materials has recently attracted considerable attentions for their catalytic,\cite{Luo2015Mar,Wrasman2018Oct} plasmonic\cite{Ramade2018Nov,Ramade2019Jun} and magnetic\cite{Capiod2019Mar,Ferrando2016Jan} properties but also features a very rich structural landscape including different chemical orderings\cite{Forster2019Oct} as well as different polymorphs\cite{Ko2018Mar}. In particular, NiAl alloys are known for high-temperature structural applications including the well-known $\gamma-\gamma$' superalloys\cite{Reed:2006th} made of nano- to micro-sized L1$_2$ \ch{Ni3Al} precipitates that exhibit very unique mechanical properties\cite{Maass2012Jun,Schloesser2013Oct,Amodeo:2017ko}.

Here, we use atomistic simulations to show that increasing the cooling rate favors the emergence of a novel metastable structure. To understand this result, we perform free energy calculations on the liquid to nanocrystal transitions using the potential energy and an entropy surrogate as structure-agnostic order parameters. Our results show that while only the most stable structure (L1$_2$) is found at room temperature,  a second free energy basin associated to an alternative structure (DO$_3$) becomes preponderant at higher temperature. Altogether, this work  further contributes to the rational control of crystal polymorph in nano-alloys which is crucial for a wide range of technological applications, at the same time enhancing our understanding of out of equilibrium nanoparticle formation through competing nucleation pathways. 

\vspace{0.5cm}
We studied nanoparticles made of $1000$ atoms. As a comparison, recent studies that focused on different materials \ch{AgPt}\cite{Pavan2015Nov}, \ch{CuPt}\cite{Rossi2018Jun} and \ch{AuAg}\cite{Gould2016Nov} considered particles of less than half this size. To initialize the calculations, atoms were randomly arranged in a sphere of radius $12.5$\,\AA\,embedded in a large simulation box ($40$\,\AA). A timestep of $0.001$\,ps was employed thorough the calculations. Then, the system was equilibrated at $2000$\,K  in the NVT ensemble. From there, the temperature was linearly decreased to 300\,K (with the same thermostat) during a time $\tau$ that was varied from $10$\,ns to $200$\,ns, corresponding to cooling rates in the range $8.5-170$\,K/ns. These values are faster than typical liquid and gas quenching \cite{Koga2006Mar} but is on par with chaotic bottom-up synthesis including laser ablation in liquids\cite{Amans2017Mar,Shih2020Apr}. Throughout the study, we employed the EAM interatomic potentials developed by Pun and Mishin which is known to correctly mimic experimental lattice and defect properties of various \ch{NiAl} systems including L1$_{2}$ \ch{Ni3Al} \citep{PurjaPun2009Dec}. In addition, this potential correctly reproduces \textit{ab initio} formation energies of binary BCC D0$_3$ and HCP D0$_{19}$ Ni${_3}$Al.  The NVT ensemble is obtained by using the Noose-Hoover thermostat with a damping factor equal to $0.1$\,ps.   Simulations were carried out using LAMMPS\cite{plimpton1995} patched with PLUMED\cite{Bonomi2019Jul,Tribello2014Feb}. Crystal analysis of the structures was performed using common neighbor analysis\cite{Honeycutt1987Sep} as implemented in OVITO\cite{Stukowski2009Dec}.


For a given cooling rate, we performed $N_{\mathrm{tot}}$ simulations using various initial conditions (positions and velocities). In the resulting nanoparticles roughly half the atoms are in the first and second surface layers and are necessarily not identified as crystallized. Therefore, a simulation is considered crystallizing when the system reaches more than 40\% of crystal-like atomic environments and the number of crystallizing simulations is denoted $N_{\mathrm{crys}}$. Otherwise, the system remains trapped in an amorphous configuration. $N_{\mathrm{tot}}$ and $N_{\mathrm{crys}}$ for the different quench durations are reported in Table\,\ref{tab:Xcrys}. The results show that as the quench duration is increased, the proportion of crystallizing simulations also increases from 0\% ($\tau$=10 ns) to more than $90$\% ($\tau$=200 ns). This is an expected result since longer quenching times allow the system to evolve under quasi-equilibrium conditions.

\begin{table}
\begin{tabular}{ p{1.5cm} p{1.5cm} p{1.5cm} p{1.5cm} p{1.5cm}}
\hline
  & \multicolumn{4}{c}{$\tau$ [ns]} \\
  & 10 & 50 & 100 &200 \\
  \cline{2-5}
  ${N_{\mathrm{tot}}}$  & 20 & 33 & 16 & 14 \\
  ${N_{\mathrm{crys}}}/{N_{\mathrm{tot}}}$ & 0 & 0.33 & 0.75 & 0.93 \\
\hline
\end{tabular}
\caption{\label{tab:Xcrys} Number of simulations ${N_{\mathrm{tot}}}$ for each investigated quench duration $\tau$. ${N_{\mathrm{crys}}}/{N_{\mathrm{tot}}}$ refers to the ratio of crystallized configurations. }
\end{table}

Before proceeding to discuss our freezing simulations, we first recall the polymorphs that have been found experimentally. In the bulk phase, \ch{Ni3Al} is known to be stable at the binary FCC L1$_2$ structure\cite{Stoloff1989Jan}. \ch{Ni3Al}-based BCC structures are less commonly observed, but can be obtained using additional alloying elements such as Fe, Ti or Ru \cite{Polvani:1976fc, Feng:2004ja}. In our simulations, several types of nanostructures are observed. They are composed of single- and polycrystals, and are including one or more polymorphs. Both binary FCC L1$_2$ and BCC D0$_3$ crystals are identified, as well as planar defects such as complex stacking faults and twin boundaries, both characterized by HCP local environments. Pseudo-twins and few HCP D0$_{19}$ nanoprecipitates are also sparingly detected which is consistent with previous experimental\cite{Kovarik:2009bj} and numerical\cite{Amodeo:2017ko} observations. Typical final configurations for $\tau=50$, $100$ and $200$\,ns along with the temporal evolution of the fraction of BCC, FCC, and HCP-like environments are shown in Fig.\,\ref{fig:MD_global} (see Supplemental Materials for all cases). The different nanocrystals can be rationalized in 3 categories (i) FCC rich, (ii) BCC rich and (iii) FCC + BCC, all with or without planar defects.

\begin{figure}
\includegraphics[width=1.\linewidth]{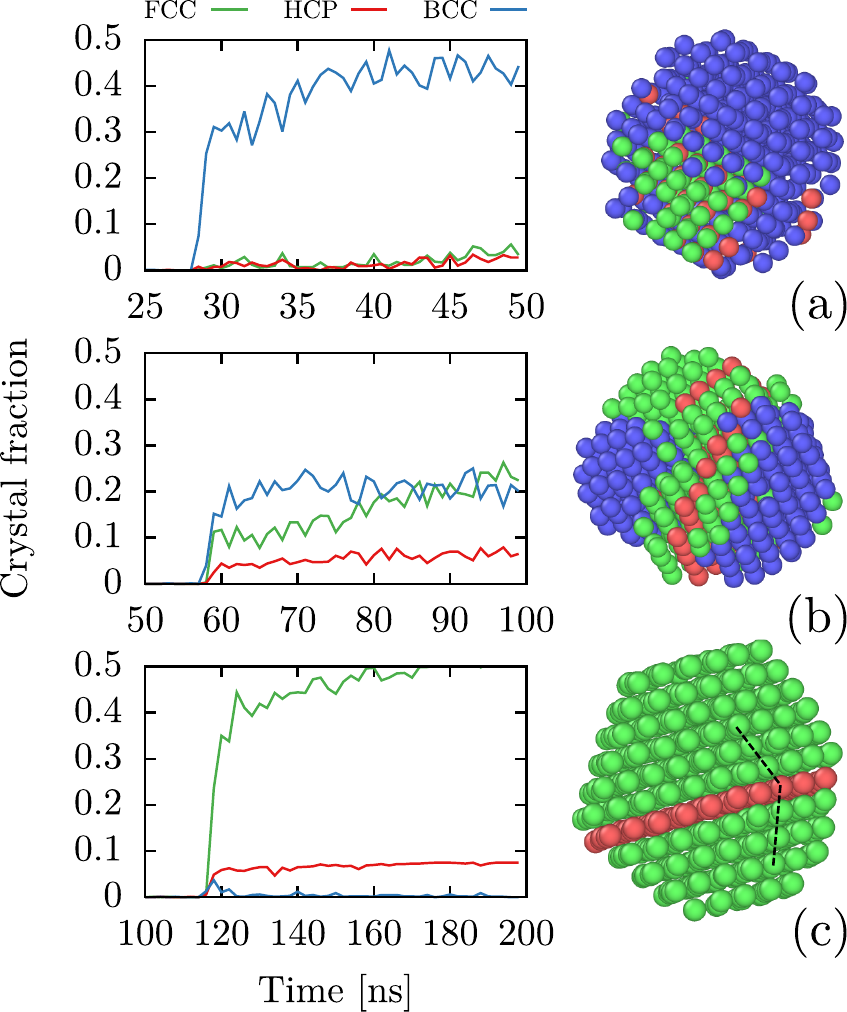}
\caption{Examples of crystallization kinetics and resulting \ch{Ni3Al} nanocrystals. Atoms color coding: Blue = BCC (D0$_3$), Red = HCP (planar defects), Green = FCC (L1$_2$) local environments. Non-crystalline atoms are removed for the sake of clarity. (a) BCC rich crystal ($\tau=50$\,ns). Only few atoms in FCC and HCP environments are identified. (b) Mixed FCC-BCC polycrystal ($\tau= 100$\,ns). (c) FCC rich crystal ($\tau=200$\,ns). The sample is a twinned bicrystal (twin boundaries formed by one HCP layer). The black dashed-line illustrates the classical mirror relation from both sides of the twin boundaries.}
\label{fig:MD_global}
\end{figure}

\begin{figure}[ht]
\includegraphics[width=1.\linewidth]{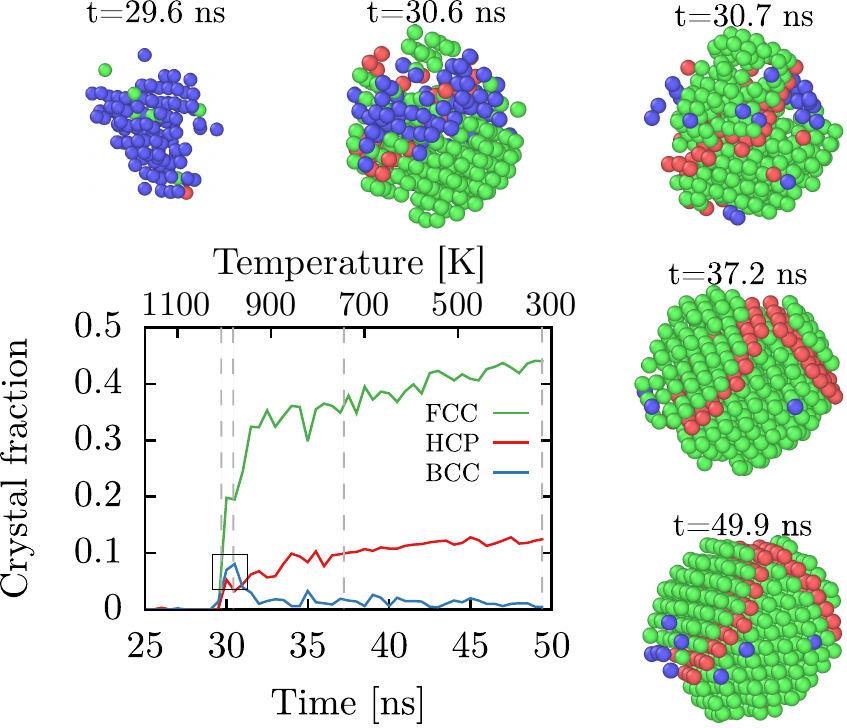}
\caption{Emergence of a BCC peak during \ch{Ni3Al} nanoparticle crystallization. Color coding corresponds to the previous figure. Grey dashed lines correspond to the snapshots. The black square emphasized the BCC peak.}
\label{fig:MD_BCCpeak}
\end{figure}

In the simulations, each crystallized nanoparticle is characterized by an initial BCC D0$_3$ nucleus, even when the final structure is FCC rich.
This feature is emphasized by a relentless BCC peak at the beginning of the liquid-to-crystal transformation, as shown for example in Fig.\,\ref{fig:MD_global}c. Then, the BCC nucleus can either grow and fill most of the particle's volume, coexist with FCC crystals or more interestingly transform into a binary FCC via a reverse martensitic transformation \cite{Krauss:1963im, Kajiwara:1979if, Kajiwara:1999gd}. This phase transformation is illustrated in Fig.\,\ref{fig:MD_BCCpeak}. In this simulation, the first liquid-to-solid transition is identified at $t=29.6$\,ns where a D0$_{3}$ BCC crystallite nucleates. Then, at a different location within the nanoparticle, the expected L1$_{2}$ FCC structure emerges and at $t=30.6$\,ns the two phases coexist. Few tens of picoseconds later, the BCC crystal transforms into an FCC one as shown at $t=37.2$\,ns. One can note that (i) the grains relative orientation and the remaining twin boundaries foster a phase transformation process rather than grain growth, (ii) most of the crystallization process happens within few tens of ns, around $T=1000$\,K. Therefore, one can speculate that most of the twin boundaries retrieved from  L1$_{2}$ nanocrystals in the simulation originate from a local BCC-to-FCC phase transformation. 

While it is the first time the binary BCC D0$_{3}$ phase of \ch{Ni_3Al} is characterized from a liquid-to-solid transformation, the phase has already been observed in MD simulations of \ch{Ni_3Al} nanowires under tensile tests using a using a slightly different interatomic potential \cite{Mishin:2004uh, Wang:2013tu}. In the latter studies, the authors observed the aforementioned transformation for particularly low diameters ($<5$\,nm) only, where the surface over volume ratio is particularly elevated, demonstrating the key-role of harmonic elasticity at the nanoscale, under extremely high stress (in the GPa range).
These results are in good agreement with our simulations and a similar argument may apply here as (i) the crystal embryo is particularly distorted at the liquid-to-solid transition and (ii) only small nanodroplets are considered. Furthermore, the presence of a binary BCC embryo at the early stage of nucleation is also consistent with results previously obtained in bulk with two simple systems made of Lennard-Jones atoms\cite{TenWolde1995Oct,Desgranges2007Jun,Lam2018Mar} and hard-spheres\cite{Auer2001Feb}. In our case, we are studying a more complex system which involves (i) nanostructures, (ii) many-body interactions, (iii) multi-component systems and (iv) quenching temperature with time but still manage to observe the emergence of the BCC structure which indicates that this process may be ubiquitous in liquid to solid transitions. 

\begin{figure}[h!]
\includegraphics[width=8.6cm]{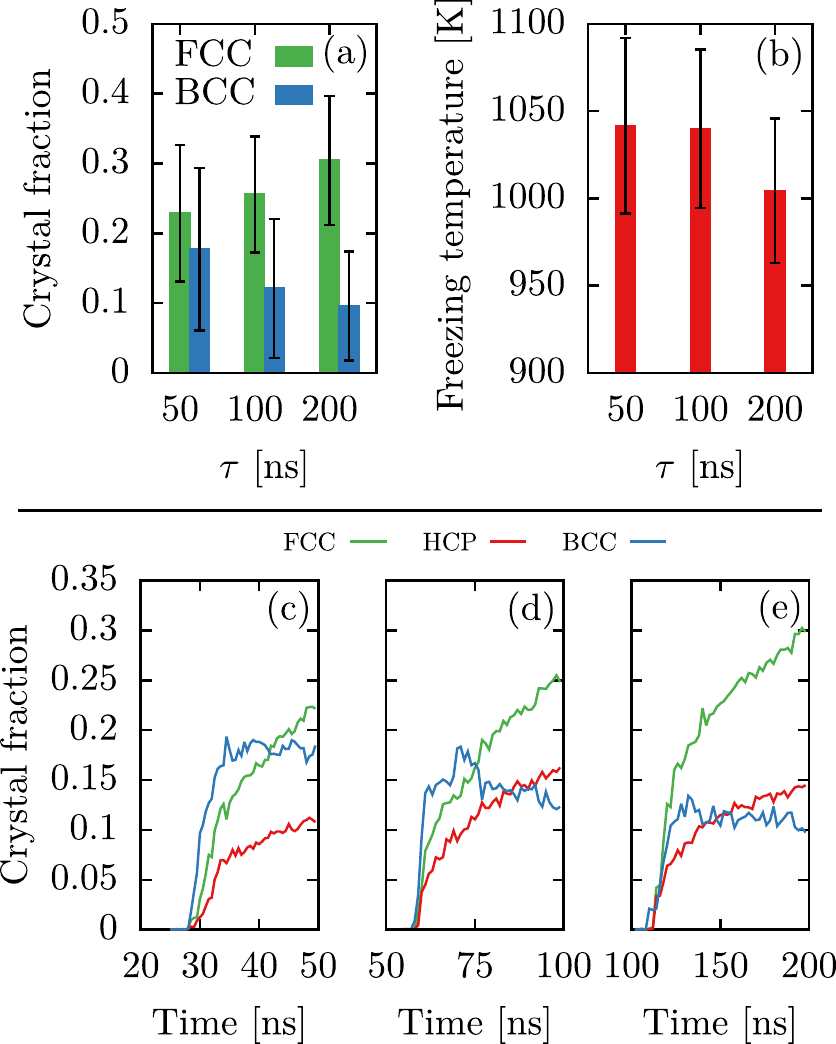}
\caption{(a,b) Fraction of the different crystal structures and  freezing temperature as a function of the quench duration $\tau$. Temporal evolution of the crystal fraction for (c) $\tau=50$\,ns, (d) $\tau=100$\,ns and (e) $\tau=200$\,ns. Results for (a-e) are averaged over all of the simulations.} 
\label{End}
\end{figure}

Our results reveal the competition between the L1$_2$ FCC and the D0$_{3}$ BCC crystal structures of \ch{Ni3Al}. These can be observed for all of the investigated values of $\tau$, but in various statistical proportions. Fig.\,\ref{End}.a shows the proportion of each crystal structure as a function of the quench duration. In general, the FCC structure is the most likely to stabilize at the end of the simulation for all the investigated values of $\tau$. However, its preponderance is decreased for the lowest quench duration while the BCC structure has the opposite behavior. In addition, for each of the simulations, we recorded the time and temperature for which an ordered cluster of $10$\,atoms is formed [See Fig.\,\ref{End}.b]. It appears that the freezing temperature is located at approximately $1000$\,K and slightly decreases with the quench duration which is typical of a thermodynamic hysteresis\cite{Rossi2018Jun}. Finally, in Fig.\,\ref{End}.(c,d,e), the temporal evolution averaged over all of the crystallizing simulations confirms that (i) BCC structures emerge at the early stages of nucleation and that (ii) decreasing the quench duration drives the system towards further stabilization of the BCC structure. We note that this averaging hides the fact that BCC can also be the most preponderant structure at the end of some of the freezing simulations [See Fig.\ref{fig:MD_global}.a]. In the following, the presence of the BCC crystallites will be investigated using metadynamics simulations. 


\vspace{0.5cm}

In order to better understand the complex mechanisms associated with the formation of the \ch{Ni_3Al} nanoparticles, we estimated the crystallization free energy landscapes using metadynamics simulations \cite{Laio02}. In this kind of simulations, the choice of the collective variable (CV) is crucial\cite{Tribello2017Mar}. In general, traditional CVs include the size of the biggest crystalline cluster\cite{Cheng2020Jan} and Steinhardt’s parameters\cite{Desgranges2019Nov}, which tend to bias the system towards a specific crystal structure. As an alternative, CVs defined from the structure of the initial and final states of the system were proposed, based on permutation invariant vectors\cite{Pipolo2017Dec} or on specific peaks of the X-ray diffractogram\cite{Zhang2019Mar,Pavan2015Nov}. In the case of polymorph selection, it becomes convenient to employ structurally agnostic collective variables to avoid any constraint on the crystal structure. For that purpose, we use the effective entropy (denoted $S$), as recently developed by Piaggi et al.\cite{Piaggi2018Oct,Piaggi2017Sep,Piaggi2017Jul}, together with the potential energy (denoted $E$). While the latter is directly computed from the force field,  $S$ is evaluated using an approximation based on liquid state theory:
\begin{equation}
S=-2\pi \rho k_B \int^\infty_0 [g(r) \ln g(r) - g(r) +1] r^2 dr
\end{equation}
where $\rho$ is the density of the system, $k_B$ is the Boltzman constant and $g(r)$ is the two-body correlation function that can be numerically computed at each simulation step\cite{Piaggi2017Jul} within a cutoff of $5$\,\AA. 

For the metadynamics simulations, the width of the deposited Gaussians along CVs $S$ and $E$ are chosen as twice the standard deviation of their distribution in the liquid regime, giving  $\delta S=0.125$\,kJ/mol/K and $\delta E=0.50$\,kJ/mol. We adopted a Gaussian height of $5$\,kJ/mol and a deposition interval of $0.1$\,ps. The free energy sampling is carried out during $80$\,ns which is similar to what was done with \ch{Ag_{147}Pt_{147}}\cite{Pavan2015Nov}. Our setup allows us to observe several reversible transitions and to achieve a satisfactory convergence of the free energy landscapes [see Supplemental Material for quantitative details]. 

\begin{figure}[t!]
\includegraphics[width=8.6cm]{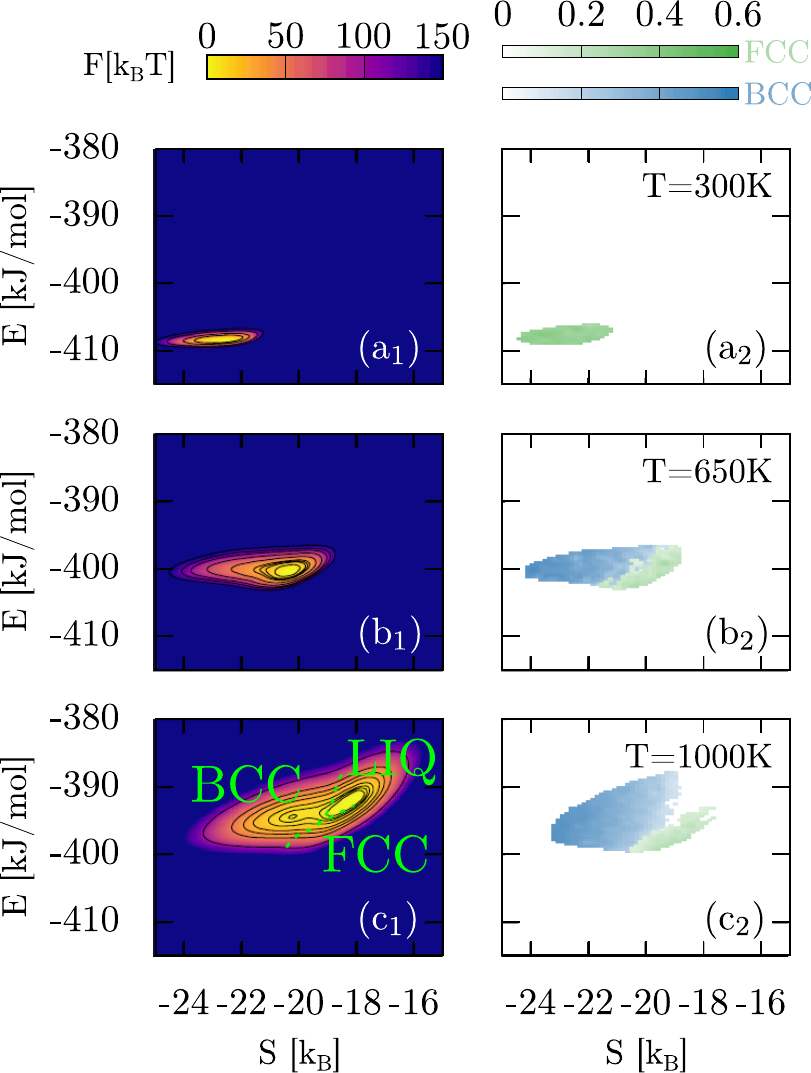}
\caption{Left panel: Free energy landscape obtained with ($S$,$E$) as collective variable at three different temperatures: T=300K (a$_\textrm{1}$), T=650K (b$_\textrm{1}$) and T=1000K (c$_\textrm{1}$). Right panel: Mapping of the average proportion of BCC and FCC structures along the metadynamic trajectory at the same temperatures. Regions with free energy larger than 150kT were not considered in this curve. The dotted green line correspond to the separation between liquid, FCC and BCC structures. }
\label{FreeEnergy}
\end{figure}

Fig.\,\ref{FreeEnergy}.a shows the free energy profile at $300$\,K along with the FCC and BCC structure fraction projected onto the ($S$,$E$) space. It appears that -- as expected for moderate temperatures -- the system adopts the L1$_2$ FCC structure. No other minimum is detected within a free-energy difference of 150\,$k_BT$. Interestingly, at $650$\,K a second region associated with the D0$_3$ BCC structure emerges, while remaining less stable than the L1$_2$ region [See Fig.\,\ref{FreeEnergy}.b]. Finally, for the largest temperature of $1000$\,K -- corresponding to the phase change temperature observed in the freezing MD simulations -- the free energy landscape becomes more complex. Indeed, a liquid region emerges in the high-energy ($E=-390$\,kJ/mol) and high-effective entropy domain ($S=-17.5\,k_B$). In addition, the crystalline region is divided in two, with the BCC region now the more stable and the FCC region metastable. 

Altogether, the metadynamics calculations show that the free energy landscape is highly dependent on the temperature with a competition between D0$_3$ BCC and L1$_2$ FCC regions. We show that the  D0$_3$ BCC region is the most stable at high temperature while the L1$_2$ FCC region becomes preponderant at room temperature. Based on these findings, we suggest the following scenario able to rationalize the formation of nanoparticles upon freezing: (1) At high temperature ($T>1000$\,K), the system remains in the liquid state; (2) At approximately $1000$\,K, a crystalline region emerges with BCC as its most favorable structure; (3) If the nucleation core appears in this temperature region, the system is more likely to grow by keeping the BCC structure, even if FCC growth is also possible; (4) If the nucleation event missed this temperature range, the systems follows the most stable scenario and grows as FCC; (5) The two types of nucleation events can occur concurrently and in this case, the system relaxes towards a coexisting structure.

\vspace{0.5cm}
As a perspective, we stress that the very surprising nucleation scenario observed here could probably be witnessed with other bimetallic nanoparticles as well as different types of nanoparticles including oxydes and chalcogenides. To study those, we foresee the use of a similar strategy combining direct freezing simulations with metadynamics calculations using the effective entropy as collective variable, as it allowed for reconstructing the nucleation process of \ch{Ni3Al} nanoparticles along different competing pathways without any a priori on the crystalline structures.

From the experimental point of view, \ch{Ni3Al} nanoparticles have mostly been produced with cooling rates that were slower or quicker than what is investigated in our simulations. Indeed, on the one hand, inert gas phase condensation\cite{Meesala2014Sep}, plasma methods\cite{Wang2006Aug,Xu2014May} and mechanical alloying\cite{Yu2010May} involved slower cooling rates and led to the formation of the most stable crystal structure ie. FCC L1$_2$. On the other hand, pulsed laser deposition with very quick cooling rates ($10^{4}$\,K/ns) produced amorphous thin films\cite{Bysakh2001May}. As such, our numerical simulations are consistent with all of these experimental results since we obtained FCC L1$_2$ and amorphous nanoparticles respectively with the slowest and the quickest simulated cooling rates. Furthermore, our findings indicate that one could stabilize the so-far unseen BCC DO$_3$ structure by adjusting the cooling rate in between these two regimes, for example by changing experimental conditions in inert gas phase condensation or in laser ablation\cite{Shih2020Apr}. 
Our work elucidates the critical issue of the competition between thermodynamics and kinetics in the out of equilibrium synthesis of nanoparticles, and it indicates a route towards the rational design and control of experimental growth processes.

\subsection*{Acknowledgement}
JL acknowledges financial support of the Fonds de la Recherche Scientifique - FNRS. Computational resources have been provided by the Consortium des Equipements de Calcul Intensif (CECI) and by the F\'ed\'eration Lyonnaise de Mod\'elisation et Sciences Numériques (FLMSN). The authors are also grateful to James F. Lutsko, David Amans, Magali Benoit, Michel Perez, Xavier Boulnat and Joel Bonneville for fruitful discussions. All authors finally thank Pablo M. Piaggi for carefully reading the manuscript and  providing insightful comments.

\providecommand{\latin}[1]{#1}
\makeatletter
\providecommand{\doi}
  {\begingroup\let\do\@makeother\dospecials
  \catcode`\{=1 \catcode`\}=2 \doi@aux}
\providecommand{\doi@aux}[1]{\endgroup\texttt{#1}}
\makeatother
\providecommand*\mcitethebibliography{\thebibliography}
\csname @ifundefined\endcsname{endmcitethebibliography}
  {\let\endmcitethebibliography\endthebibliography}{}

\end{document}